\documentclass[preprint,superscriptaddress,nobibnotes]{revtex4}

\usepackage{times}
\usepackage[pdftex]{graphicx}
\usepackage[pdftex]{epsfig}
\usepackage{amsmath}
\usepackage{pslatex}
\usepackage{amssymb}
\usepackage{amsfonts}
\usepackage{color}
\usepackage{wrapfig}
\usepackage{eucal}
\usepackage{hhline}
\usepackage{threeparttable}
\usepackage{supertabular}
\usepackage{multirow}
\usepackage{tabularx}

\newcolumntype{Y}{>{\centering\arraybackslash}X}

\begin{document}

\title{A multi-spectral and polarization-selective surface-plasmon resonant mid-infrared detector}

 \author{Jessie Rosenberg}
 \affiliation{Thomas J. Watson, Sr.,\ Laboratory of Applied Physics, California Institute of Technology, Pasadena, CA 91125, USA}
 \author{Rajeev V. Shenoi}
 \affiliation{Center for High Technology Materials (ECE Dept), University of New Mexico, Albuquerque, NM 87106}
 \author{Thomas E. Vandervelde}
 \affiliation{Center for High Technology Materials (ECE Dept), University of New Mexico, Albuquerque, NM 87106}
 \author{Sanjay Krishna}
 \email{skrishna@chtm.unm.edu}
 \affiliation{Center for High Technology Materials (ECE Dept), University of New Mexico, Albuquerque, NM 87106}
 \author{Oskar Painter}
 \email{opainter@caltech.edu}
 \homepage{http://copilot.caltech.edu}
 \affiliation{Thomas J. Watson, Sr.,\ Laboratory of Applied Physics, California Institute of Technology, Pasadena, CA 91125, USA}
 \date{\today}

\date{\today}


\begin{abstract}
We demonstrate a multi-spectral polarization sensitive mid-infrared dots-in-a-well (DWELL) photodetector utilizing surface-plasmonic resonant elements, with tailorable frequency response and polarization selectivity.  The resonant responsivity of the surface-plasmon detector shows an enhancement of up to 5 times that of an unpatterned control detector.  As the plasmonic resonator involves only surface patterning of the top metal contact, this method is independent of light-absorbing material and can easily be integrated with current focal plane array processing for imaging applications.  
\end{abstract}

\maketitle

\setcounter{page}{1}

\noindent


\noindent Multicolor mid-infrared imaging is a highly desirable technology in a number of applications such as night vision, missile tracking, medical diagnostics, and environmental monitoring~\cite{Rogalski:2008}. The current dominant technologies in the field of multispectral imaging rely on the use of either a broadband focal plane array (FPA) with a spinning filter wheel in front of it~\cite{Althouse:1991}, or a bank of FPAs with a dispersive element such as a grating or prism to separate light of different frequencies. These methods are limited by the often high cost and complexity of such systems. However, if spectral sensitivity could be encoded at the pixel level within a single focal plane array, multi-spectral detection would become much more practical for use in a wide range of applications. In addition, the use of pixel-integrated resonators to provide spectral sensitivity can dramatically increase the efficiency of the detector due to the many passes light makes within the resonator.

The most common methods for pixel-level encoding utilize epitaxially defined dielectric Fabry Perot cavities~\cite{Huang:2004, Han:2005, Attaluri:2007} or lithographically patterned gratings~\cite{Goossen:1985, Hasnain:1989, Andersson:May1991}, but these can be difficult to integrate into standard FPA processing. Previously pixel-integrated resonant detectors utilizing dielectric photonic crystals (PCs)~\cite{Posani:2006, Yang:2008}, deep-etched plasmonic PCs~\cite{Shenoi:2007}, and metallic Fabry Perot cavities~\cite{Hu:2008} have been shown in the mid-infrared region, as well as plasmonic color-selective gratings in the visible region~\cite{Laux:2008, White:2009}. Here we design and demonstrate mid-infrared frequency- and polarization-specific plasmonic photonic crystal detector pixels in dots-in-a-well (DWELL) photodetector material~\cite{Krishna:PhysD2005, Shenoi:2008} utilizing a shallow etch design which extends only through the top plasmonic metal layer. This resonator design can be used with any detector material system, does not introduce any etching damage to the detector active material, and adds only one lithography step to current FPA processing techniques. Finite element modeling successfully reproduces the experimentally measured spectral response, and also shows good agreement with the measured level of responsivity enhancement at the plasmonic cavity resonant wavelengths.

\begin{figure*}
\begin{center}
\includegraphics[width=\columnwidth]{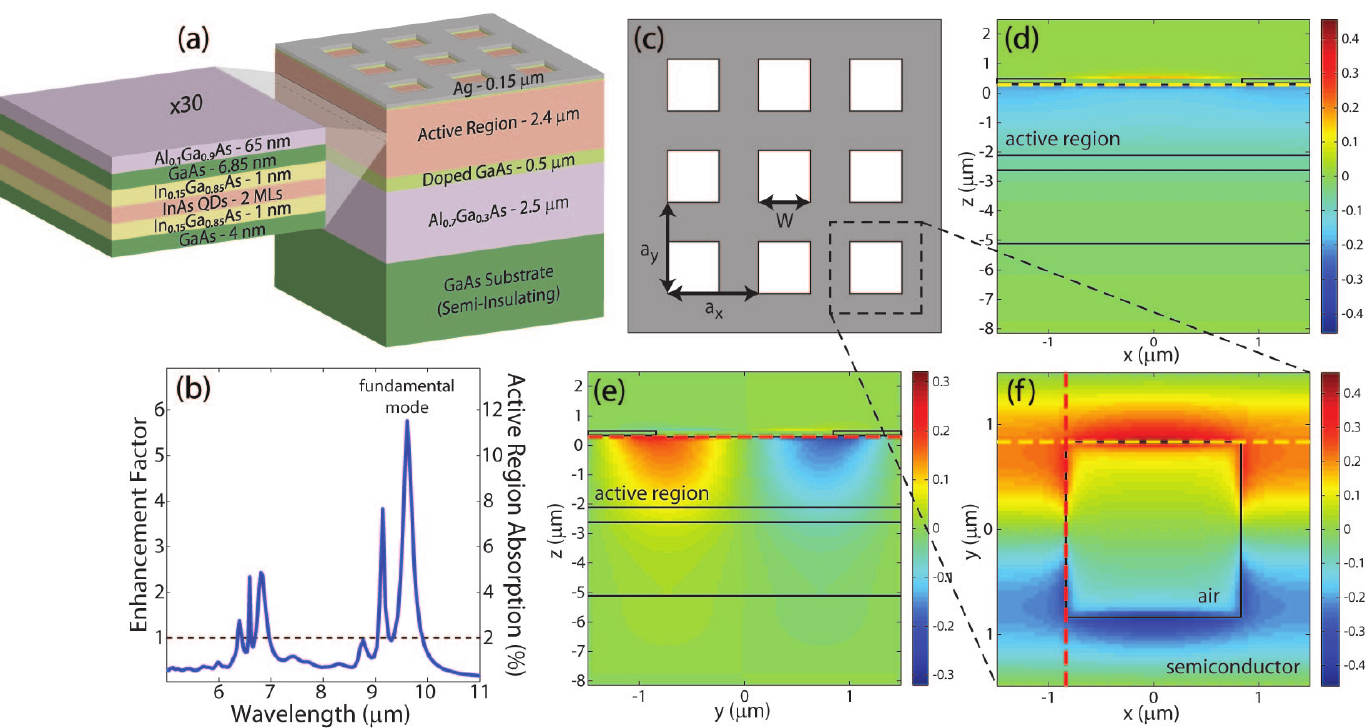}
\caption{(a) A cross-section of the device structure, showing the composition of the dots-in-a-well (DWELL) detector material. (b) FDTD simulated enhancement factor and active region absorption versus wavelength, based on a 2\% single-pass absorption. All simulations use a structure with lattice constant $a=2.939$~$\mu$m, $\bar{W}=0.567$, and metal thickness $t=150$~nm. (c) A diagram of the unstretched PC structure showing relevant dimensions. The expanded plots show the $E_z$ mode profile for one lattice constant of one of the two dipole modes for an unstretched PC lattice in (d) the $x$-$z$ plane along the hole edge, (e) the $y$-$z$ plane along the hole edge, and (f) the $x$-$y$ plane just beneath the metal-semiconductor interface.}
 \label{fig:dipole_modes}
\end{center}
\end{figure*}

The resonant cavity used in this work consists of a single layer of metal with etched square holes in a square lattice periodic array. A representation of several lattice constants of the device structure is shown in Fig.~\ref{fig:dipole_modes}(a). The plasmonic layer provides the vertical confinement, confining the optical mode with a maximum at the surface of the metal (Fig.~\ref{fig:dipole_modes}(d,e)), while the etched air-holes create a PC pattern to confine the light in-plane. Combined together, this resonator design provides full 3D confinement, significantly increasing the amount of time light spends within the detector active region, and therefore enhancing the probability of detection. Due to the strong index contrast between the surface plasmon~\cite{Raether:1988, Prade:1991, Barnes:2003} mode beneath the metal regions and the dielectric-confined mode beneath the air holes, this plasmonic PC grating is strong enough to generate an in-plane confined resonant mode without etching into the detector active material~\cite{Pendry:2004, Bahriz:2007}, allowing a resonator to be fabricated without damaging or removing active material. A detailed numerical and symmetry analysis~\cite{ref:Jessie2} shows that the two degenerate dipole-like in-plane modes of the structure (Fig.~\ref{fig:dipole_modes}(d-f)) couple most easily from free space. Further improvements in the free-space coupling efficiency were performed by optimizing the top metal thickness and hole size. In addition, as the two dipole-like modes couple to orthogonal polarizations of incoming light, a stretch of the PC lattice breaks the degeneracy of the two modes, splitting their resonance frequencies and thus achieving high polarization selectivity~\cite{Painter:2002, Yang:2008}.

The finite difference time domain (FDTD) simulated active region absorption and enhancement factor (details of which are presented in Ref.~\onlinecite{ref:Jessie2}) are plotted versus frequency in Fig.~\ref{fig:dipole_modes}(b) for a perfectly periodic structure, showing the fundamental plasmonic (degenerate) dipole-like modes at $\lambda=9.6$~$\mu$m (the two shorter wavelength resonance peaks correspond to higher-order vertical waveguide modes of the DWELL epitaxy).  The series of resonance peaks near $\lambda=6.5$~$\mu$m correspond to higher-order modes of the in-plane square lattice PC.  Figure~\ref{fig:dipole_modes}(c) shows relevant dimensions of the structure; we define the normalized hole width as $\bar{W}=2W/(a_x + a_y)$. Field profiles along different planes for the square lattice dipole-like mode are plotted in Fig.~\ref{fig:dipole_modes}(d-f) over a single unit cell; the other dipole mode has the same field pattern, rotated by 90 degrees in the $x$-$y$ plane. The simulated active region absorption corresponding to this mode is 11.5\%, given an approximate 2\% single-pass absorption in the DWELL material, and corresponds to an expected resonant responsivity enhancement of 5-6 times that of a control sample with no plasmonic layer or PC patterning.

\begin{figure}
\centering
\includegraphics[width=0.75\columnwidth]{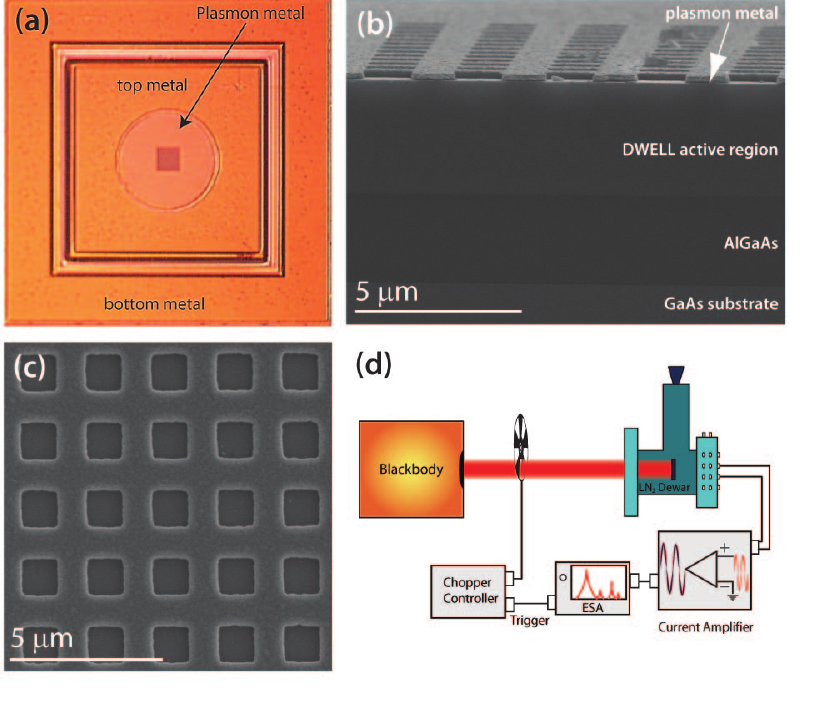}
\caption{(a) Optical image of the fabricated device indicating the top, bottom and plasmon metallizations. (b) Cross-sectional SEM of the fabricated square-lattice device indicating the plasmon metal and device layers. (c) SEM image of the fabricated square-lattice PC pattern on the plasmon metallization. (d) Schematic of the setup used for measuring responsivity and detectivity of the fabricated devices. Responsivity measurements were performed by illuminating the sample with a calibrated Mikron M365 blackbody at $T=800$~K. The blackbody radiation was modulated at a frequency of 400~Hz using a chopper and this signal was used as a trigger for the SRS 760 fast Fourier transform (FFT) spectrum analyzer. The photocurrent was amplified using a SRS 570 low noise amplifier and then measured in the spectrum analyzer.}
\label{fig:device_images}
\end{figure}

\begin{figure*}
\centering
\includegraphics[width=\columnwidth]{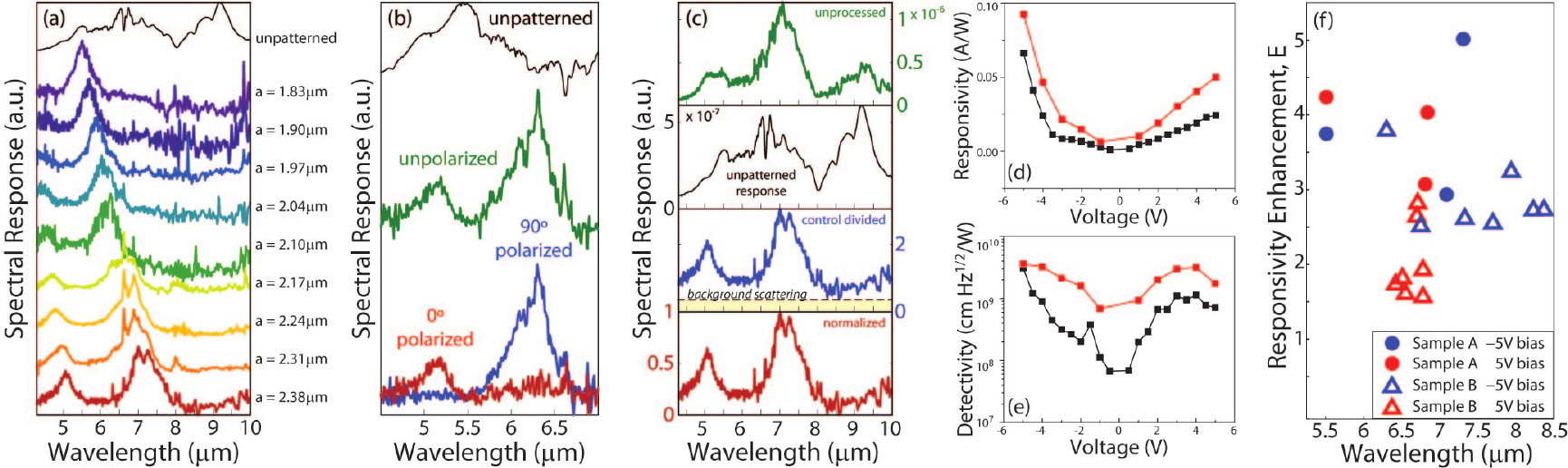}
\caption{(a) Normalized spectral response from square lattice devices at a bias of 5~V, indicating tuning of peak wavelength with the lattice constant. (b) Normalized spectral response from rectangular lattice devices at a bias of -5~V. The response to unpolarized light is shown in green, and beneath is the response to light polarized at 0 degrees (red) and 90 degrees (blue) relative to the shorter lattice constant dimension of the lattice. (c) Data processing of the $a = 2.38~\mu$m device response. The unprocessed data (green) is divided by the unpatterned DWELL response to show the resonances independently of the base detector response. The background scattering from other regions of the sample (yellow) is then subtracted from the control divided data (blue) and normalized, to give the final spectral response (red) that is plotted in Fig. \ref{fig:responsivity}(a,b). (d) Measured responsivity and (e) detectivity of a stretched lattice device with $a_x = 2.164~\mu$m and $a_y = 1.776~\mu$m (red), and a control device (black). (f) Measured peak responsivity enhancement $E$ versus resonant device wavelength for two samples at positive and negative bias.}
\label{fig:responsivity}
\end{figure*}

To test these predictions, we fabricated two detector samples: sample $A$ with a square lattice PC, and sample $B$ with a rectangular lattice PC having a lattice constant stretching ratio $a_y/a_x=1.2$. Figure~\ref{fig:device_images}(a-c) shows representative images of the fabricated devices. All of the PC patterns had the same normalized hole width ($\bar{W}$), but different lattice constant values, and therefore different resonant wavelengths determined by the scaling of the pattern. The spectral response of the surface-plasmon resonant detectors was measured using a Nicolet 870 Fourier transform infrared spectrometer (FTIR) with the patterned detector sample used in place of the standard FTIR detector. Responsivity and detectivity measurements were performed using the experimental setup shown in Fig.~\ref{fig:device_images}(d). To separate out the background scattering and the frequency response of the DWELL material from the resonant enhancement, we perform data processing as shown for a typical spectral response measurement in Fig.~\ref{fig:responsivity}(c).

As predicted, by varying the lattice constant and symmetry of the patterned grating, we are able to tailor the wavelength and polarization response of each detector pixel. Figure~\ref{fig:responsivity}(a) shows the resonant spectral response from a set of representative detector pixels on sample $A$, varying the peak wavelength response from $5.5$~$\mu$m to $7.2$~$\mu$m by choosing PC lattice constants in a range from $1.82$~$\mu$m to $2.45$~$\mu$m. The linewidth of these resonances is $\sim 0.9~\mu$m, providing strong spectral sensitivity within the broad background DWELL response which covers more than 5~$\mu$m. In addition to the fundamental surface plasmon resonant mode, we also observe a higher-order plasmon mode as predicted by the FDTD simulations in Fig.~\ref{fig:dipole_modes}(b), at a wavelength in good agreement with the theory. In order to generate a polarization-sensitive response, we stretch the lattice constant in one direction (sample $B$), splitting the resonant detector response into two well-separated peaks as shown in the green curve of Fig.~\ref{fig:responsivity}(b). By varying the polarization of the light incident on the detector, we show that these two peaks correspond to orthogonal linear polarization directions of incoming light, as represented by the blue and red curves of Fig.~\ref{fig:responsivity}(b). The high polarization extinction between the two curves indicates clearly the strong polarization dependence in our device. The experimentally measured spectral peaks are broadened relative to the FDTD simulated values in Fig.~\ref{fig:dipole_modes}(b) due to the finite extent of the PC pattern ($\sim$50~$\mu$m in diameter), and therefore the limited in-plane confinement.

To characterize the efficiency of the detector response and the resonant enhancement, we define and measure the responsivity and detectivity of samples $A$ and $B$ as follows. The peak responsivity was computed using the expression 
\begin{equation}\label{responsivity2}
R_p=\frac{I_0}{{\int_{\lambda_{1}}^{\lambda_2}R_{N}(\lambda)L_e(\lambda,T)A_sA_d\frac{tF_F}{r^2}d\lambda}}
\end{equation}
\noindent where $R_{N}(\lambda)$, $I_0$, $L_e$, $T$, $A_s$, and $A_d $ are the normalized spectral response, measured photocurrent, the black body spectral excitance, the black body source temperature, the area of the source, and the area of the detector, and $r$, $t$, $F_F$ are the distance between the source and the detector, the transmission of the window and geometrical form factor, respectively. The lower and upper wavelength bounds of the detector response are given by $\lambda_1$ and $\lambda_2$. The detectivity $D^*$ is then 
\begin{equation}\label{dstar}
D^*=\frac{\sqrt{A_d \Delta f}}{i_n}R_p
\end{equation}
\noindent where  $A_d$ is the detector area, $\Delta f$ is the noise equivalent bandwidth ofour measurement, and $i_n$ is the noise current.

Compared to a control (unpatterned) sample, the plasmonic PC patterned devices provide a strong enhancement of responsivity and a corresponding increase in detectivity across an applied bias range from -5~V to 5~V, as shown in Fig.~\ref{fig:responsivity}(d) and (e). The enhancement factor $E$ is defined as $E=R(\lambda_i)/R_{c}(\lambda_i)$, where $ R(\lambda_i)$ is the responsivity of the patterned detector at the resonant wavelength and $R_{c}(\lambda_i)$ is the responsivity of the control sample at the same wavelength. In Fig.~\ref{fig:responsivity}(f), we show enhancement factors across a range of wavelengths reaching as high as 5X for sample $A$, and 4X for sample $B$. With further improvement to the free-space coupling of the resonators and the cavity quality factor (allowing for critical coupling of the incoming radiation~\cite{ref:Cai}) through incorporation of a bottom metal layer forming a double-metal cavity\cite{ref:Jessie2}, we estimate this factor could be improved by another factor of 5.

In summary, we have designed and fabricated a multi-spectral mid-infrared photodetector utilizing a plasmonic resonator for frequency sensitivity, polarization splitting, and responsivity enhancement. We demonstrate control over the peak wavelength response of the detector pixels, strong polarization selectivity, and resonant responsivity enhancements of up to 5X. Though we chose the valuable mid-infrared region for this demonstration, the resonator design can be extended for use at wavelengths from the terahertz to the visible with suitable scaling of the PC holes. As it involves only a surface patterning, it can be used with various light absorbing materials and detector designs. The single-metal waveguide structure shown here can also be extended to a double-metal structure that would readily integrate with current FPA processing, and would enable higher quality factor modes with better active region confinement and thus higher frequency selectivity and responsivity enhancement.

The authors would like to thank R. Perahia for help in the initial device processing.  This work was supported by the AFOSR through grant \#FA9550-06-1-0443 and MURI grant \#FA9550-04-1-0434, the AFRL through grant \#FA9453-07-C-0171, and the IC post-doctoral program.


\begin{thebibliography}{10}

\bibitem{Rogalski:2008}
A.~Rogalski,
\newblock Opto-Electronics Review {\bf 16}, 458 (2008).

\bibitem{Althouse:1991}
M.~L. Althouse and C.~I. Chang,
\newblock Optical Engineering {\bf 30}, 1725 (1991).

\bibitem{Huang:2004}
H.~Huang, Y.~Huang, X.~Wang, Q.~Wang, and X.~Ren,
\newblock Photonics Technology Letters, IEEE {\bf 16}, 245 (2004).

\bibitem{Han:2005}
Q.~Han et~al.,
\newblock Applied Physics Letters {\bf 87}, 111105+ (2005).

\bibitem{Attaluri:2007}
R.~S. Attaluri et~al.,
\newblock Journal of Vacuum Science \& Technology B: Microelectronics and
  Nanometer Structures {\bf 25}, 1186 (2007).

\bibitem{Goossen:1985}
K.~W. Goossen and S.~A. Lyon,
\newblock Applied Physics Letters {\bf 47}, 1257 (1985).

\bibitem{Hasnain:1989}
G.~Hasnain et~al.,
\newblock Applied Physics Letters {\bf 54}, 2515 (1989).

\bibitem{Andersson:May1991}
J.~Y. Andersson, L.~Lundqvist, and Z.~F. Paska,
\newblock Applied Physics Letters {\bf 58}, 2264 (1991).

\bibitem{Posani:2006}
K.~T. Posani et~al.,
\newblock Applied Physics Letters {\bf 88}, 151104+ (2006).

\bibitem{Yang:2008}
J.~K. Yang, M.~K. Seo, I.~K. Hwang, S.~B. Kim, and Y.~H. Lee,
\newblock Applied Physics Letters {\bf 93}, 211103+ (2008).

\bibitem{Shenoi:2007}
R.~V. Shenoi et~al.,
\newblock Plasmon assisted photonic crystal quantum dot sensors,
\newblock volume 6713, pages 67130P+, SPIE, 2007.

\bibitem{Hu:2008}
X.~Hu et~al.,
\newblock Applied Physics Letters {\bf 93}, 241108+ (2008).

\bibitem{Laux:2008}
E.~Laux, C.~Genet, T.~Skauli, and T.~W. Ebbesen,
\newblock Nat Photon {\bf 2}, 161 (2008).

\bibitem{White:2009}
J.~S. White et~al.,
\newblock Opt. Lett. {\bf 34}, 686 (2009).

\bibitem{Krishna:PhysD2005}
S.~Krishna,
\newblock Journal of Physics D: Applied Physics {\bf 38}, 2142 (2005).

\bibitem{Shenoi:2008}
R.~V. Shenoi et~al.,
\newblock Journal of Vacuum Science and Technology B {\bf 26}, 1136 (2008).

\bibitem{Raether:1988}
H.~Raether,
\newblock {\em Surface Plasmons on Smooth and Rough Surfaces and on Gratings
  (Springer Tracts in Modern Physics)},
\newblock Springer, 1988.

\bibitem{Prade:1991}
B.~Prade, J.~Y. Vinet, and A.~Mysyrowicz,
\newblock Phys. Rev. B {\bf 44}, 13556 (1991).

\bibitem{Barnes:2003}
W.~L. Barnes, A.~Dereux, and T.~W. Ebbesen,
\newblock Nature {\bf 424}, 824 (2003).

\bibitem{Pendry:2004}
J.~B. Pendry, L.~Martin-Moreno, and F.~J. Garcia-Vidal,
\newblock Science {\bf 305}, 847 (2004).

\bibitem{Bahriz:2007}
M.~Bahriz, V.~Moreau, R.~Colombelli, O.~Crisafulli, and O.~Painter,
\newblock Optics Express {\bf 15}, 5948 (2007).

\bibitem{ref:Jessie2}
J.~Rosenberg, R.~V. Shenoi, S.~Krishna, and O.~Painter,
\newblock (2009), manuscript in preparation.

\bibitem{Painter:2002}
O.~Painter and K.~Srinivasan,
\newblock Optics Letters {\bf 27}, 339 (2002).

\bibitem{ref:Cai}
M.~Cai, O.~Painter, and K.~J. Vahala,
\newblock Phys. Rev. Lett. {\bf 85}, 74 (2000).

\end{thebibliography}

\bibliographystyle{osa}

\end{document}